\documentclass[10pt,a4paper]{article}
\usepackage[utf8]{inputenc}
\usepackage{amsmath}
\usepackage{amsfonts}
\usepackage{amssymb}
\usepackage{graphicx}

\newif\ifshowfigures

\showfigurestrue 

\title{An Early Model of Transistors and Circuits}

\author{{ \normalsize L. da F. Costa$^1$, Filipi N. Silva$^1$, Cesar H. Comin$^1$}}

\date{\emph{{\small $^1$S\~ao Carlos Institute of Physics, University of S\~ao Paulo, PO Box 369, 13560-970, S\~ao Carlos, SP, Brazil }}\\[3mm] {\small \today}}

\begin{document}

\maketitle

\begin{abstract}
Bipolar junction transistors (BJTs) have been at the core of linear electronics from its beginnings. Although their properties can be well represented transport model equations, design and analysis approaches have, to a good extent, been limited to using current-equispaced horizontal and parallel isolines of the characteristic surface. Here, we resort to the geometrical structure imposed on BJTs behavior by the Early effect and voltage as a means to derive a simple, intuitive and more complete respective model that, though excluding cut-off and saturation regimes, can simplify the design and characterization of BJTs and respective circuits, including integrated devices. The approach involves using a beam of isolines converging at the Early voltage as the model of a BJT. The angles of the isolines, another important aspect of BJTs, are experimentally verified often to vary almost in linear fashion with the base current, endowing the proposed Early model with an intrinsic geometric nature. An experimental-numerical methodology is suggested for the Early model estimation, and applied to 12 real-world small signal NPN BJTs. Interesting results are obtained, including the identification of diverse Early voltage values for different BJT types and the observation, for the considered BJTs, of the almost linear variation of the isoline angles with the input current. A case example of transfer function estimation, as well as total harmonic distortion analysis, are also provided in order to illustrate the potential of the proposed methodology for approximating, under certain circumstances, the operation of the common emitter configuration.  The proposed model also has good potential as a didactic aid in physics and electronics programs, complementing the currently adopted modeling approaches.
\end{abstract}

\section{Introduction}

Despite (or because) of its simplicity, linearity constitutes a pot of gold in science and technology.  Indeed, a great deal of modeling in science has been based on some type of linearization of the problem around an operating point, so that the involved differential equations remain linear and amenable of analytical solution.  In linear systems, linearity is sought of because it \emph{does not affect} the shape of the signals, except for a multiplying factor $\alpha$, which is used to control and modify the signal magnitude.   For instance, $\alpha > 1$ will amplify the signal, giving rise to an \emph{amplifier}.  Other values of $\alpha$ are also useful in respective situations, while linearity is also a sought property for other purposes such as filters and linear superimposition.  Interestingly, despite its mathematical simplicity, linearity is particularly hard to find or enforce in nature, as virtually every real-world device/system will present some kind of intrinsic non-linearity at specific time and  spatial scales.  Such a fact has implied great efforts by researchers and practitioners in order to find system configurations capable of providing suitable linearity~\cite{tsividis1986mos,wang2004capacitance,braid2linearity}.

One area that has witnessed great activity regarding the search for linearity is \emph{linear electronics}.  Here, the goal is to obtain amplifiers, mixers and filters that are as linear as possible for a wide range of practical applications.  The non-linearity in an electronic amplifier normally starts at the very adopted devices (e.g. vacuum tubes and discrete or integrated solid state devices).  As mentioned above, all such real-world components have intrinsic non-linearities of diverse natures, which will conspire to deform the signals~\cite{self2013audio,cordell2011designing}.  So, a naturally good starting point to tackle the linearity problem is to better understand the non-linearity intrinsic to such devices.   Fortunately, this can be effectively achieved by  developing a model of the devices, capable of capturing their non-linearities preferably in a way that is  simple, objective and effective. Backed by experimental validations, such models can be of great help for understanding and devising methods capable of improving linearity.  Here, we focus attention on transistors, more specifically bipolar types (BJTs), and their use in discrete or integrated circuits.

A natural candidate for transistor model are its respective transport equations, derived from semiconductor physics and, particularly, the hybrid models (e.g.\cite{jaeger1997microelectronic}).  However, such an approach is seemingly complicated, involving several parameters and the exponential function weighted by thermodynamical parameters.  In practice, it is common to simplify the transistor behavior in the linear region in terms of current-equispaced horizontal parallel lines (in both collector current $I_c$ and base current $I_b$) placed in the $current \times voltage$ device space.  Such an approach appears in many textbooks both as a didactic aid and as a possible design methodology.  When transistors are used in circuits, such as in the common emitter configuration shown in Figure~\ref{f:amplifier}, a particularly interesting type of model of the circuit operation can be obtained based on the concept of transfer function~\cite{laughton2013electrical,silva2016seeking}, because such a representation defines a graphical relationship between input and output signals, which is particularly suitable for inferring the linearity of the respective device/configuration.  

The Early effect, and related \emph{Early voltage}, were identified by James M. Early in 1952~\cite{early1952effects}.  In BJTs, the Early effect is characterized by an increase of collector current as a consequence of collector voltage augmentation.  In the physical device, this occurs because of a narrowing of the base width as implied by extensions of the base-collector depletion region~\cite{jaeger1997microelectronic,yuan1991improved}. 
The Early voltage, which appears in some transport model-derived equations and is particularly  related to the hybrid pi model, is often left out in practice.  Deriving from semiconductor physics, such a voltage ultimately implies that the virtual prolongation of the isolines in the characteristic surface of a device tend to intersect at a value $-V_a$ along non-positive portion of the voltage device axis.   In a previous work~\cite{costa2016negative}, the Early voltage was used in order to select specific isolines of BJTs for negative feedback analysis.  That same work also suggested that small signal BJTs from the same type tend to have  characteristic properties, enough to discriminate them into several respective clusters, and that negative feedback may not provide fully invariance to such device differences.  That work also identified that the properties of several real-world BJTs tend to be statistically explained by 1 or 2 degrees of freedom, suggesting an intrinsic simplicity to BJT behavior, which is explained in the present work.

Because of its geometrical simplicity, and its ability to provide a transistor model that is more complete than the parallel isolines that have been often considered, the Early voltage organization provides a natural principle for obtaining a simple, yet effective, geometrical model of BJT operation.  This is precisely the main objective of the present work.  After reviewing the basic concepts of common emitter configuration, characteristic surfaces, and transfer functions, we proceed to presenting the Early approach, starting from a more complete version of BJT equation, incorporating the Early effect.  For simplicity's sake, the model assumes as hypothesis that the angles of the isolines vary linearly in terms of the base current, a hypothesis that has been to a good extent corroborated experimentally in this work respectively to 12 real-world small signal NPN BJTs.  The Early approach is also used to obtain a simple yet effective model of the common-emitter configuration, allowing the respective transfer function to be immediately obtained.  So, we address the Early modeling at two successive levels:  one applicable to BJT devices, and another aimed at BJT circuits such as the common emitter configuration.   We derive the respective voltage and current transfer functions along specific load lines, and propose a geometrical definition of the BJT properties of current gain, output resistance and transresistance.  Implications of such Early model to the linearity of amplifiers in the common emitter configuration are then presented and discussed in theoretical fashion, as well as in terms of THD analysis of a real-world device.  The concept of \emph{local} Early modeling of a real circuit is also motivated and developed.

This article also describes a numeric-computational procedure for estimating the Early voltage from a set of experimental isolines obtained from real-world BJTs.  This procedure involves linear regression along specific portions of interest of the isolines, followed by normalization and characterization of the sharpness of the $V_a$ intersections.  A measurement of how linear the isoline angles vary with the base current is also provided.  These methods are then used to investigate a set of 12 real-world NPN BJTs operating in class A common emitter configuration.  A series of interesting results are obtained, including the tendency of the isoline angles to vary in almost linear fashion with the base current, and the identification of distinct values of $V_a$ for different transistor types.  A case example illustrating how the proposed Early model can assist the design and characterization of amplifiers is also provided and discussed, as well as total harmonic distortion analysis (THD) of transfer functions derived by the Early approach.  Ultimately, the Early voltage by itself seems to determine most of the properties of the considered BJTs, which has interesting practical and theoretical implications.

\section{Basic Concepts}

\begin{figure}[]
  \begin{center}
  \includegraphics[width=0.4\linewidth]{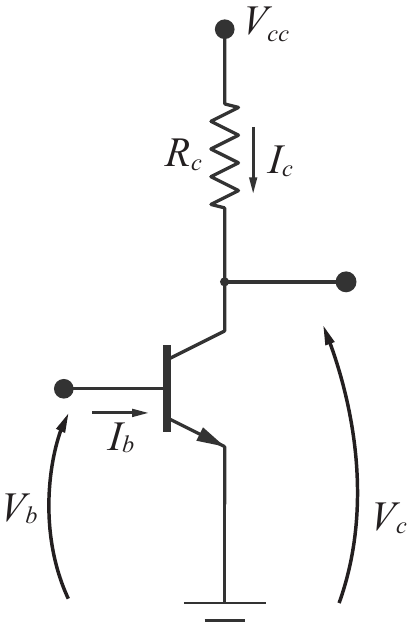} \\
  \caption{Common emitter configuration of a BJT.  The input signal is represented by $I_b$, and the output corresponds to $I_c$ or $V_c$. The base-emitter junction is directly biased, while the base-collector junction is reversely biased.  Resistor $R_c$ is often understood as the load of the circuit. $V_{cc}$ is the fixed voltage for circuit operation.}
  ~\label{f:amplifier}
  \end{center}
\end{figure}

Studying and modeling BJTs involve estimating and mathematically/geometrically interrelating their parameters and properties.  In the present work, the common emitter configuration~\cite{self2013audio}, illustrated in Figure~\ref{f:amplifier}, is henceforth adopted in order to illustrate the derivation of such parameters and properties by using the proposed BJT Early model.

The resistor $R_c$ is connected to the collector of the NPN BJT, giving rise to a circuit that includes both input (the base) and output (collector) ports.  The \emph{common emitter} nomenclature derives from the fact that the BJT emitter is shared by both input and output meshes, being henceforth directly (as opposed to using an emitter resistance) connected to ground in order to  minimize feedback, as we want to study the intrinsic BJT properties.  Observe that a fixed voltage $V_{cc}$ is imposed upon $R_c$, but no current will flow through this resistance unless some positive base current $I_b$ is supplied.  Typically,  $I_b$ and $I_c$ (or $V_c$) are understood as the input and output of the circuit, respectively.

In the common emitter configuration, the base emitter junction is directly biased, while the base collector junction is reversely biased, which allows $I_c$ to be controlled by $I_b$ or $V_b$.  As $I_b$ is increased, more current $I_c$ will flow through $R_c$, reducing the voltage $V_c$ at the BJT collector.  In BJTs the \emph{device} current gain, $\beta$, is typically larger than one (typically, in small signal BJTs, it varies from 20 to 300), hence the amplifier operation.  In order to cater for the amplification of both negative and positive portions of the input signal $I_b(t)$, this circuit is normally biased so that some current $I_c>0$ flows at the operation point $\Omega = (V_{co},I_{bo})$ even for zero dynamical input ($I_b=0$), which is typical of class A operation (e.g. ~\cite{self2013audio}).

In case the imaginary reactances of a BJT can be neglected (e.g. when working at relatively low frequencies compared to the BJT time constants), the electrical behavior of a BJT can be captured by its characteristic surface $I_b =  f(I_c, V_c)$ (e.g.~\cite{silva2016seeking}).  Because this surface, including its boundary behavior, is intrinsic to each specific device and therefore can vary from case to case, they have to be derived experimentally. Actually, the present work is mainly motivated at obtaining a simple, yet effective, model that can be used to  approximate this surface.   Most parameters and electrical properties of the BJT can be derived from the characteristic surface.  For instance, the current gain $\beta$ corresponds to 

\begin{equation}
  \beta = \frac{\partial I_c}{\partial I_b}
\end{equation}

Other parameters of interest include the BJT output resistance $R_o$ and its transresistance $R_T$, given respectively as

\begin{eqnarray}
  R_o = \frac{\partial V_c}{\partial I_c} \\
  R_T = \frac{\partial V_c}{\partial I_b}
\end{eqnarray} 

Of particular interest, because of its impact on the linearity (and therefore quality) of the amplification, are the \emph{transfer functions} (e.g. \cite{silva2016seeking}) derived from the surface $I_b =  f(I_c, V_c)$ and the voltage and current constraints imposed by $V_{cc}$ and $R_c$.  It follows immediately from circuit analysis of the common emitter configuration in Figure~\ref{f:amplifier} that  $V_c = V_{cc} - R_c I_c$, which corresponds to a respective \emph{load line} of the circuit.  Because both $I_c$ and $V_c$ are parameterized by $I_b$ --- i.e.\ $I_c = I_c(I_b)$ and $V_c = V_c(I_b)$, we actually have that 

\begin{equation}
  I_c(I_b) = V_{cc} - R_c I_c(I_b)
\end{equation}

Recall that $I_b$ is, by its turn, related with $V_b$.  The current and voltage \emph{transfer functions} can be obtained by expressing either $I_c$ or $V_c$ in terms of $I_b$.  Observe the role of $V_{cc}$ and $R_c$ in defining such transfer functions by slicing the surface $I_b =  f(I_c, V_c)$ along a specific and respective straight line restriction.  Though, ideally, the transfer functions should correspond to perfectly straight lines, in practice this is never achieved as a consequence of intrinsic non-linearities of real-world devices.  Negative feedback constitutes a traditional means to improve the transfer function linearity.

\subsection{Total Harmonic Distortion}

A traditional approach to quantify the non-linearity of transfer functions is the total harmonic distortion (THD)\cite{cordell2011designing,raikos2009low}. The THD is calculated by considering as input to the system a pure sinusoidal function with frequency $f$, and measuring the magnitude of the spurious harmonics that are generated at the system's output. Specifically, representing as $V_f$ the magnitude of the fundamental and as $V_{2f}$, $V_{3f}$, $V_{4f}$, \dots the magnitude of the spurious frequencies, the THD is calculated as

\begin{equation}
THD(f) = {\sqrt{V_{2f}^2 + V_{3f}^2 + V_{4f}^2 + \cdots} \over V_f}.\label{eq:THDDef}
\end{equation}
If the system is only composed of purely resistive loads, the same THD will be attained for any input frequency $f$.

\section{Device and Circuit Early Voltage Modeling}

Now, we are in a position to proceed to developing the Early model for expressing the geometry of the BJT characteristic surface, and then extending it to incorporate the effect of $I_b$, assuming purely resistive operation (or relatively low frequencies).  It should be observed that the current approach is not intended to account for saturation and cut-off regimes.  We also derive the current and voltage transfer functions implied by the common emitter configuration.

We start with the BJT equation, relating $I_c$ in terms of $V_b$~(e.g. \cite{jaeger1997microelectronic})

\begin{equation}
  I_c = I_s \, e^{\frac{V_{b}}{V_T}} \left(1 + \frac{V_{c}}{V_a} \right) 
\end{equation}
where $I_c$ is the collector current, $I_s$ is the reverse saturation current (characteristic of each specific device), $V_{b}$ is the base voltage, $V_{c}$ is the collector voltage, $V_T$ is the thermal voltage ($kT/q$, in the order of $26\,mV$ at $300K$), and $V_a$ is the Early voltage (normally varying between 50V and 100V).  Observe that, for a given situation, $I_s$, $V_a$ and $V_T$ can be understood as constants, or parameters.  According to the above equation, the collector current $I_c$ is modulated by $V_b$ (typically it is possible to consider $V_b \approx r_o I_b$, where $r_o$ has small value).  For a fixed $\hat{I}_b$, we have an implied $\hat{V}_b$, so that $c = I_s e^{\frac{\hat{V}_b}{V_T}}$ becomes a constant and the above equation can be rewritten as

\begin{equation}
  I_c = c \, \left( 1 + V_c/V_a \right) \label{eq:IcXVc}
\end{equation}

\begin{figure}[]
  \begin{center}
  \includegraphics[width=\linewidth]{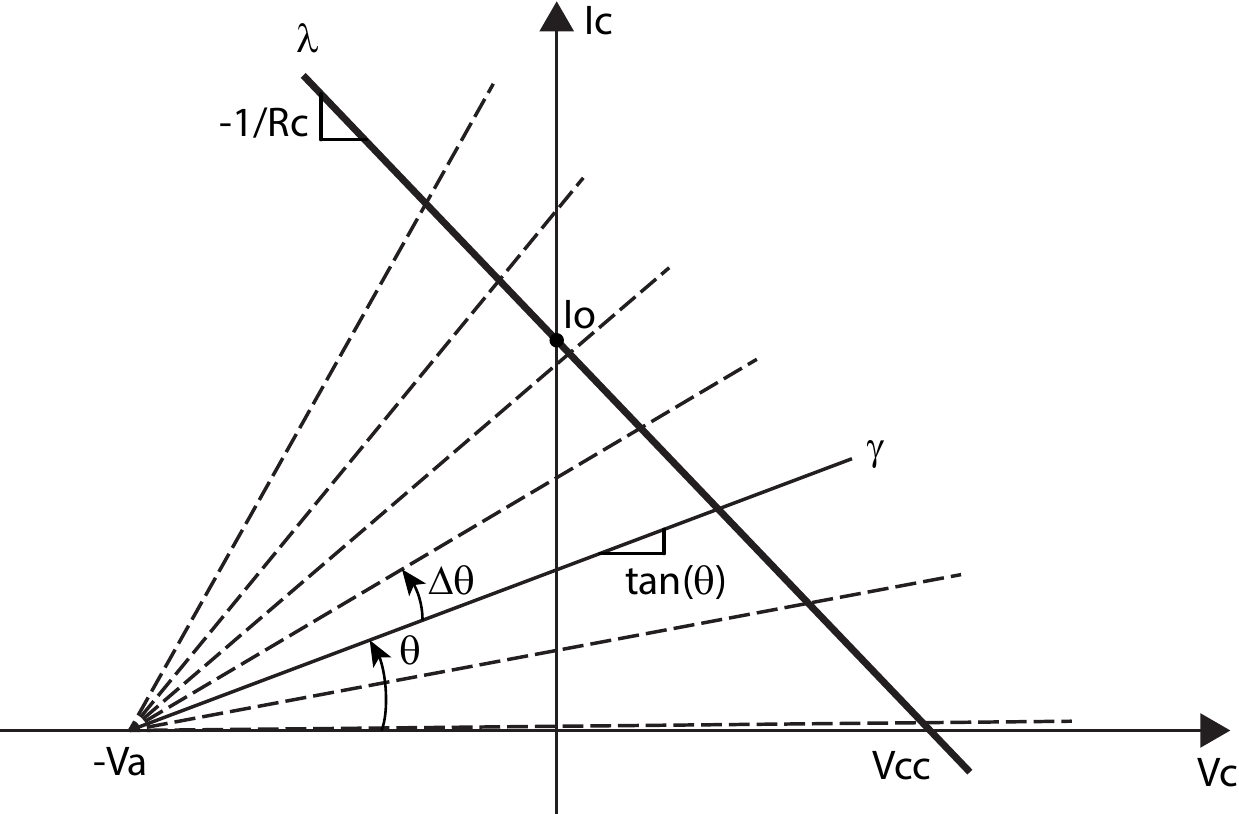} \\
  \caption{The geometrical construction used for the Early voltage model of BJTs.  The isolines are approximated by a beam of straight lines that intersect at the same value $-V_a$ along the $V_c$ axis.  The angles of the isolines, $\theta$, are assumed to vary linearly with $I_b$ which, however, does not yield device linearity.  The load line, determined by $R_c$ and $V_c$ is also depicted, representing a restriction (slicing) of the isolines, which determines the respective transfer functions. }
  ~\label{f:geometry}
  \end{center}
\end{figure}

In particular, when $I_c=0$, we have that $V_c = -V_a$, implying that all lines defined by the above equation cross the $V_c$ axis at $-V_a$.  Figure~\ref{f:geometry} shows the visualization of Equation~\ref{eq:IcXVc} for several values of $V_b$ for the specific case when the lines are spaced by the same angle.     

The proposed Early model of BJTs can be geometrically derived in the $I_c \times V_c$ space (please refer to Figure~\ref{f:geometry}).  The basic assumption is that all isolines $\gamma$ intersect one another at the point $(-V_a,0)$ along the $V_c$ axis.  Typical electronic operation is restricted to the region $I_c\geq 0$ and $V_c \geq 0$ (i.e. the first quadrant).  According to this setting, each isoline $\gamma$, indexed by the angle $\theta$, can be immediately expressed as 

\begin{equation}
   \gamma:  \:  I_c = (V_c+V_a) \, \tan(\theta)    \label{eq:gamma}
\end{equation}

Observe that this equation has $V_a$ as sole parameter (later another parameter, $s$, will be added), while $\theta$ acts as a free variable, which will be later related to the base current $I_b$.  Now, we use this model to characterize a common emitter circuit, therefore considering that the operation of the modeled BJT is now restricted by the purely resistive load line $\lambda$, given as

\begin{equation}
   \lambda: \:  I_c = (V_{cc} - V_c)/R_c    \label{eq:lambda} 
\end{equation}
This line, which is completely defined by only two parameters, the load resistance $R_c$ and the fixed power voltage $V_{cc}$, extends through the first quadrant of $I_c \times V_c$, crossing the $V_c$ axis at $Vcc$ and the $I_c$ axis at $Io = V_{cc}/R_c$.  The slope of $\lambda$ is $-1/R_c$.  

The imposition of the $\gamma$ restriction on the BJT isolines allows either $I_c$ or $V_c$ to be expressed in terms of $\theta$, yielding two respective  parametric curves that correspond to current and voltage \emph{transfer functions}.  This can be achieved by solving the system

\begin{eqnarray}
\left\{ 
   \begin{array}{l}
   I_c = (V_c + V_a) \, \tan(\theta)  \\
   I_c = (V_{cc} - V_c)/R_c
   \end{array}
\right.
\end{eqnarray}
which yields

\begin{equation}
   V_c(\theta) = \frac{V_{cc} - R_c V_a \tan(\theta)}{R_c \tan(\theta)+1}  \label{eq:Vcth}
\end{equation}
Plugging into the load line (Equation~\ref{eq:lambda}), we have

\begin{equation}
   I_c(\theta) = \frac{\left( V_{cc} + V_a \right) \, \tan(\theta) }{R_c \tan(\theta) + 1} \label{eq:Icth}
\end{equation}
The range $[0, \theta_{\max}]$ of $\theta$ values corresponding to the full extension of a load lined given by $V_{cc}$ and $R_c$ is therefore obtained as

\begin{equation}
  \theta_{\max} = \arctan(V_{cc}/(R_c \, V_a))  \label{eq:thetamax}
\end{equation}

The two transfer functions respectively given by Equations~\ref{eq:Vcth} and~\ref{eq:Icth},  derived from the proposed Early model representation of BJT operation in common emitter configuration, provide valuable information about the linearity of the device and circuit.  By taking into account only three parameters --- namely $V_a$, $V_{cc}$ and $R_c$ --- these simple equations are potentially capable of approximating the geometrical operation of a BJT in such a configuration, except for  the saturation and cut-off regions, eventual presence of imaginary reactances and the influence of the base current $I_b$, which is incorporated next.  Observe that among the three involved parameters, $V_a$ derives from the device, while $V_{cc}$ and $R_c$ are defined by the common emitter circuit configuration.  It is important to observe that the Early representation without the load line restriction can be understood as a geometrical \emph{model of the BJT itself}, while the incorporation of the load line and derivation of the respective transfer functions corresponds to a \emph{model of the common emitter circuit}.

The Early representation allows the henceforth defined \emph{geometrical} BJT parameters (or constants) to be simply derived as 

\begin{align}
  \beta_\theta & = \frac{\partial I_c}{\partial \theta} = (V_c + V_a) \, \sec^2(\theta) \\
  R_{o\theta} & = \frac{\partial V_c}{\partial I_c} = \tan^{-1}(\theta) \\
  R_{T\theta} & = \frac{\partial V_c}{\partial \theta} = -\frac{I_c}{\sin^2(\theta)}  
\end{align}

So far, we have replaced the controlling effect of $I_b$ in terms of the geometrical $\theta$ parameters.  However, it is possible to consider a more general functional relationship $\theta = g(I_b)$ linking $I_b$ to $\theta$, so as to incorporate the effect of the input base current $I_b$ on the variation of $I_c$ and $V_c$.  Geometrically, such a relationship governs how the isolines rotate as $I_b$ increases.  Any functional relationship $g()$ can be considered in the proposed model, directly affecting the linearity of the respective transfer function.  The reference situation, which is assumed in the proposed model and verified for the considered real-world BJTs, arises when $\theta$ is directly proportional to $I_b$, i.e.

\begin{equation}
  \theta(I_b) = s \,I_b    \label{eq:alpha}
\end{equation}

In such a case, if $I_b$ is varied at fixed steps $\Delta I_b$, the isolines will rotate with a constant angular displacement $\Delta \theta$, as illustrated in Figure~\ref{f:geometry}. This means that the constants of the transistor can be written as

\begin{align}
  \beta & = s\beta_\theta = s(V_c+V_a)\sec^2(sI_b) \\
  R_o & = R_{o\theta} = \tan^{-1}(sI_b) \\
  R_T & = s R_{T\theta} = -\frac{sI_c}{\sin^2(sI_b)}
\end{align} 

It is interesting to observe that, when $\theta=sI_b$, a way to achieve a linear transfer function would be to have load lines $\lambda$ corresponding to arcs of circles centered at $(-V_a, 0)$.  Alternatively, linearity will asymptotically increase by displacing the load line to become infinitely distant from the point $(-V_a,0)$.  As could be expected, when $V_a \rightarrow \infty$, the isolines become parallel one another, so that any load line will yield perfect linearity.  A third approach to linearity would correspond to adjust $g(I_b)$ to have varying speed so as to compensate for the nonuniform variation of $I_c$ implied by the uniform angle variation.

\section{Some Implications of the BJT Early Model}

\begin{figure}[]
  \begin{center}
  \includegraphics[width=0.9\linewidth]{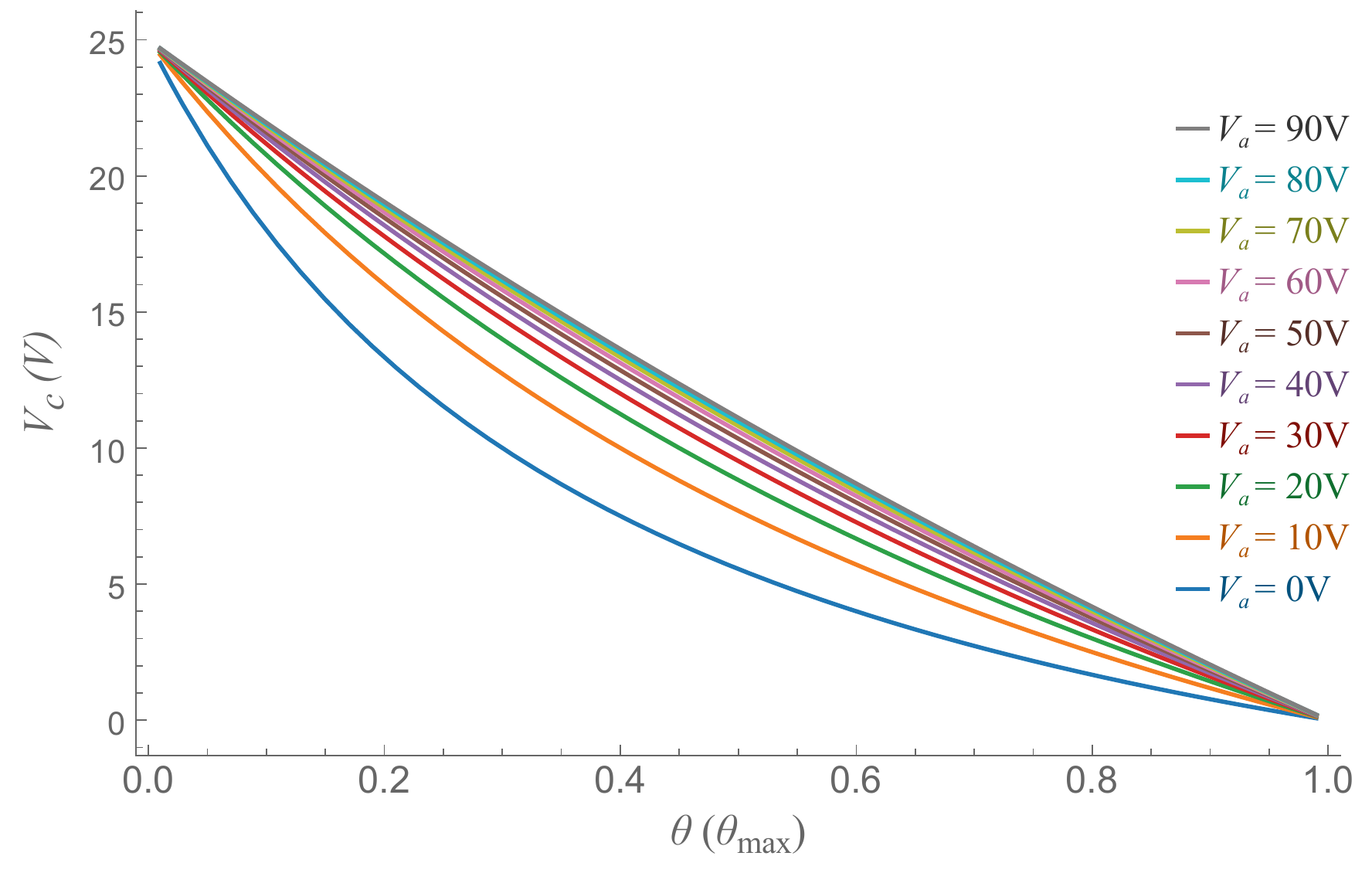} \\
  \caption{Voltage transfer functions $V_c(\theta)$ obtained by the Early model for $R_c= 900\Omega$, $V_{cc} = 25V$ and $s = 15707.96\,^o/A$.  The linearity improves asymptotically for $V_a \rightarrow \infty$.  The transfer functions obtained for the more realistic values of $V_a$ between $50V$ and $100V$ are characterized by intense non-linearity.}
  ~\label{f:transfers}
  \end{center}
\end{figure}

A number of interesting and especially critical situations can be discussed in a geometrical and conceptually simple way with the help of the suggested model.  For instance, going back to the just mentioned problem of maximizing the linearity of the transfer functions, it is straightforward to verify from the geometry of the Early model that this can be achieved by placing the operation point $\Omega$ as further as possible from the point $(-V_a,0)$.  Figure~\ref{f:transfers} shows $V_c \times I_b$ transfer functions derived through the proposed model, for several Early voltage values and the realistic setting $R_c= 900\Omega$, $V_{cc} = 25V$ and $s = 15707.96\,^o/A$ (as experimentally derived by linear regression of $\theta$ in terms of $I_b$).  It is clear from these results that the linearity of the voltage transfer function is strongly affected by $V_a$, with the best linearity being achieved for the largest value $V_a = 100V$ in this example.

It can be similarly shown that, for a selected or imposed $V_{cc}$, linearity can be improved by reducing $R_c$.  Thus, since circuit operation is restricted to the first quadrant of the $I_c \times V_c$ space, placing $\Omega$ on the point $(V_{cc},0)$ and using a very small value of $R_c$ represents a potentially optimal configuration regarding linearity (not considering the relationship $\theta \times I_b$).   Because such a situation cannot account for the negative values of the input signal, it can only be used in systems where the load is coupled to the amplifier through a transformer, since the latter can, under certain circumstances, complement the amplification for the negative amplified values as a consequence of the stored electromagnetic field.  To any extent, moving the operation point far away from $(-V_a,0)$ will necessarily increase the output resistance $R_o$, which is influenced by $\tan(\theta)$ (e.g. when considering that an output load is attached to $V_c$), limiting the applications to loads with small resistances.

\section{Local Early Model and a Respective Numerical Procedure}

\begin{figure}[]
  \begin{center}
  \includegraphics[width=\linewidth]{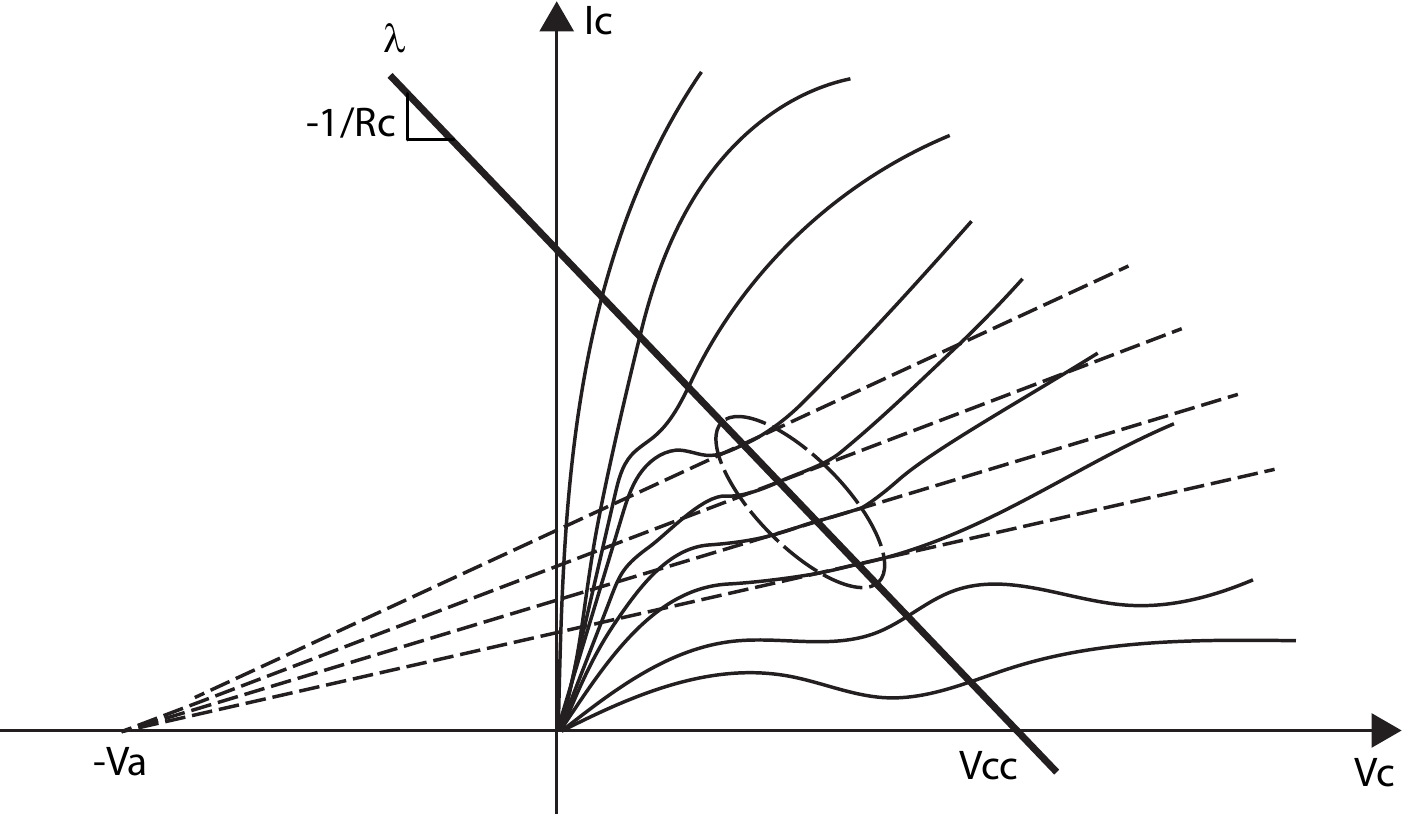} \\
  \caption{The concept of \emph{local} Early model of a BJT.  Several isolines of a hypothetical device in the space $I_c \times V_c$, generally exhibiting divergent behavior but still presenting a small region where tangent prolongations intersect one another along the $V_c$ axis.  Provided device operation is constrained to such a region (e.g. by imposing a load line, through a circuit), the proposed Early model can be extended to such situations. }
  ~\label{f:local_Early}
  \end{center}
\end{figure}

The above formulation assumes the behavior of the BJT to be geometrically describable by the Early effect.  However, situations may arise in which the convergence of isolines is verified only locally, inside a specific region of interest in the operation space $I_c \times V_c$.  This is illustrated in Figure~\ref{f:local_Early}, with respect to a hypothetical situation.  In this case, though the isolines are mostly erratic, there is a region, marked by the ellipse, where the prolongation of the tangents will still define an intersection on the $V_c$ axis.  In principle, such a region could have varied shapes, but we henceforth focus attention on straight regions, because this corresponds to the geometry implied by purely resistive loads.  In such situations, the Early representation can be immediately applied and used even for the analysis of the devices/circuits not globally following the transistor equation, provided the analysis is constrained to respective regions of interest where relatively sharp intersections can be obtained.  Interestingly, the concept of local Early model potentially paves the way for eventually finding more than on e Early voltage for a same BJT.

In contrast to the \emph{global} approach developed in the previous sections, the region-based modeling is henceforth called \emph{local} Early modeling of the BJT.  One particularly interesting question regarding real-world devices concerns to what an extent, ultimately, their behavior can be globally, or locally, modeled by the Early representation.  This important question provides the subject for the remainder of this work.   

\begin{figure}[]
  \begin{center}
  \includegraphics[width=\linewidth]{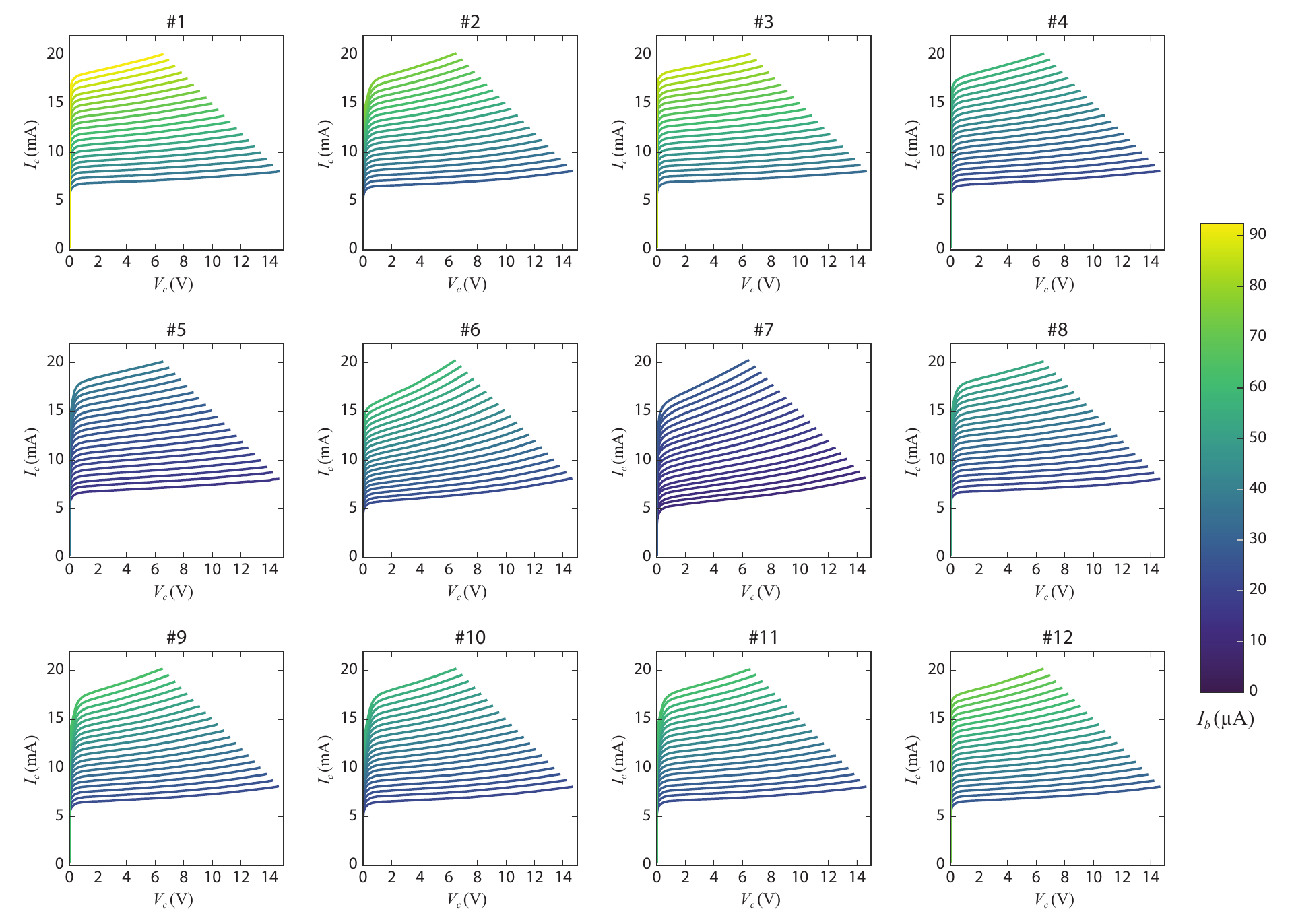} \\
  \caption{Isolines obtained for 12 real-world small signal NPN BJTs (one of each type), identified as $\#1$, $\#2$, etc.  These standardized isolines correspond to 21 equispaced $I_c$ values comprised between 8 and 20mA.  Observe the varying inclinations and shapes of the isolines for different devices. }
  ~\label{f:12_trans}
  \end{center}
\end{figure}

An experimental-numeric procedure is described next for estimation of the Early Voltage $V_a$ of a real-world BJT, leading to its respective Early model.  First, the voltage-current properties of the investigated BJT need to be experimentally estimated by using some data capture system.  In this work, this was done by using a customized microcontrolled acquisition system with simultaneous sample and holding of all monitored voltage signals. For simplicity's sake, we henceforth assume that the BJT experimental data correspond to isolines obtained for various $I_b$ values.  Such a kind of data is illustrated in Figure~\ref{f:12_trans}, which present experimentally obtained isolines respective to 12 real-world small signal NPN BJTs.

Given such experimental data, it is necessary to apply some numeric-com- putational methodology so as to estimate $V_a$, as well as the sharpness of the intersections, which is henceforth expressed as $\delta$. If the isolines were perfectly straight line segments, $V_a$ could be easily obtained by calculating the intersection, along the $V_c$ axis, of least mean square linear regressions of the isolines.  Because the isolines will never be perfectly straight (e.g.\ as a consequence of the cut-off and saturation regions, not to mention intrinsic irregularities along their extension), it is necessary to devise robust methodologies for estimating $V_a$ and also providing some quantification of the sharpness of the intersections.  In principle, there are two main approaches that can be considered: (i) choose a suitable region in the space $I_c \times V_c$ and apply linear regression on the portions of the isolines constrained by that region; and (ii) search for a region constraining the isolines so that the sharpest intersection is achieved.  Clearly, approach (ii) corresponds to an optimization methodology and is, therefore, more demanding and complex. 

In the present work, we adopted a hybrid approach where methodology (i) is applied considering several regions where to perform linear regression on  the isolines.  More specifically, first a suitable value of $R_c=\hat{R}_c$ is obtained (yielding more constant values of $V_a$ for each transistor), and then the isolines are scanned through respective load lines with fixed $\hat{R}_c$ and varying $V_{cc}$ along the $I_c \times V_c$ space.  Methodology (i) is applied to each of these henceforth called \emph{scanning load lines}.  We proceed by presenting the methodology (i) in more detail.

First, the region $R$ of the $I_c \times V_c$ space where the isolines will be considered for linear regression has to specified.  In practice, this can be done by considered an imposed load line $\lambda$, or through a design approach involving some inspection capable of identifying a promising region (characterized by isolines with converging slopes).  For instance, going back to the situation depicted in Figure~\ref{f:local_Early}, one would like to chose the region inside the ellipse.  

Once the region $R$ has been chosen, linear regression~\cite{bevington2003data} can be applied to each of the isolines, yielding a respective goodness of fit as well as the slope $m$ and intercept $c$ of the adjusted, prolonged, straight line, i.e.

\begin{equation}
   I_c = m V_c + c
\end{equation}

In this work, we will select the points within a window of width $W$ centered at the crossing of each isoline with the imposed load line. So, the intersections of the straight lines obtained through linear regression with the $V_c$ axis can be immediately obtained as $V_i = -c/m$.  Now, given a series of factors such as non-linearities of isolines, experimental error, and quantization noise, it is rather unlikely that all isolines will intersect perfectly at the same point along the $V_c$ axis.  This bears two important implications: (a) it is important to estimate the most likely value of $V_a$; and (b) it is necessary to have some quantitative indication of the sharpness of the intersection $\delta$.  Indeed, the latter value will provide a quantification of how well the Early model adheres to the real BJT in the region $R$.  Though other approaches are possible, henceforth we take $V_a$ to correspond to the mean of the intersections values $V_i$, while expressing the sharpness of the intersections in terms of the coefficient of variation of those $V_i$ values.

\begin{figure}[]
  \begin{center}
  \includegraphics[width=0.7\linewidth]{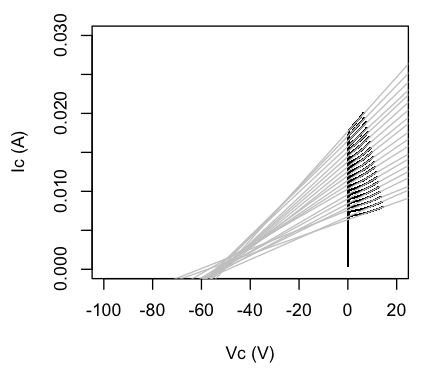} \\
  \caption{Example of numeric estimation of $V_a$, sharpness of intersections $\delta$, and rms error $\varepsilon$.  Straight lines are adjusted to portions of the isolines considering $W=23$ points centered at the intersection  of each isoline with the load line.  The intersections of such straight lines define the Early voltage $V_a$. }
  ~\label{f:intersections}
  \end{center}
\end{figure}

The above methodology is now illustrated with respect to the experimental isolines obtained for transistor \#1 (shown in Figure~\ref{f:12_trans}).  Let the region of interest  correspond to $W=23$ points centered at the crossing of each isoline (256 point-long) with the load line. Linear regression is now applied, yielding the prolongations shown in Figure~\ref{f:intersections}.  The estimated $V_a$ is $53.05V$.  Despite some minor divergence between the isolines, the overall intersection is well-characterized, with a coefficient of variation $\delta = 5.15$.  The rms error for linear relationship between $\theta$ and $I_b$ was found to be $\varepsilon = 0.019$.

\section{Experimental Results}

Provided with the numerical approach described in the previous section, we can now address some interesting implied questions, such as:  To what an extent do real BJTs adhere to the Early model, as expressed by the sharpness of intersections?  Is this verified globally or locally?  How do different types of BJTs compare one another regarding their estimated $V_a$ and sharpness of intersections?  How well can the Early model predict transfer functions?

Henceforth, we apply the proposed procedure to a lot of 12 different types of real-world small signal NPN BJTs, each represented by a respective sample. This is so because we verified in a previous work~\cite{costa2016negative} that small signal transistors from a same type tend to exhibit consistent features.  Each device/type is henceforth indicated by the $\#$ prefix.  As before, we consider $W=23$ points along each isoline, centered at the respective intersection with the considered load line, as defining the region of interest $R$.  The suggested procedure is then applied and the respective distributions obtained so that $V_a$ and the sharpness of the intersections can be estimated.  We performed an initial screening to verify how the estimated $V_a$ values behaved with $R_c$, and verified some variations.   So, we chose the configuration leading to minimal variation, for each BJT, of $V_a$ in terms of $V_{cc}$.  For the considered BJTs, this corresponds to setting $R_c \approx 800 \Omega$, which is henceforth adopted.

\begin{figure}[]
  \begin{center}
  \includegraphics[width=\linewidth]{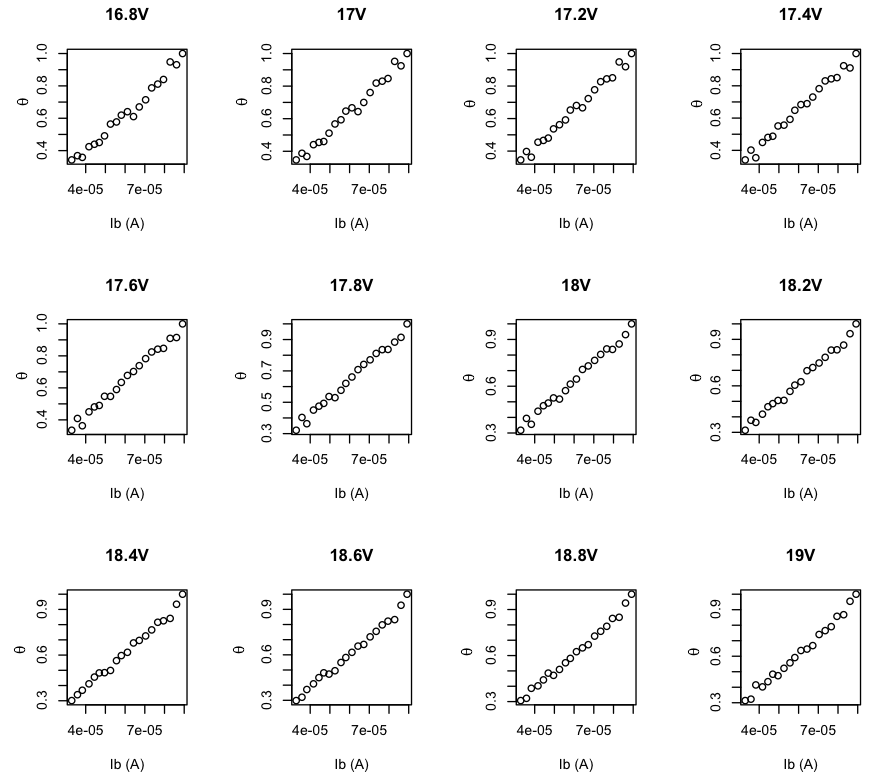} \\
  \caption{Example of how $\theta$ tends to vary with $I_b$ for transistor $\#1$, in terms of the load lines defined by successive values of $V_c$.  An almost linear relationship is observed, especially for the intermediate and large values of $V_c$. The title of each plot corresponds to the $V_{cc}$ defining the respective scanning load line used for $V_a$ estimation.}
  ~\label{f:speeds}
  \end{center}
\end{figure}

The first important issue to be investigated is how $\theta$ varies with $I_b$.  In Figure~\ref{f:speeds} we present an example of such a relationship for the isolines of transistor $\#1$, for the several considered values of $V_c$ (i.e. respective load lines). It is clear, especially for the intermediate and large values of $V_c$, that $\theta$ is almost linearly related to $I_b$.  Similar results were obtained for the other considered BJTs.

\begin{figure}[]
  \begin{center}
  \includegraphics[width=\linewidth]{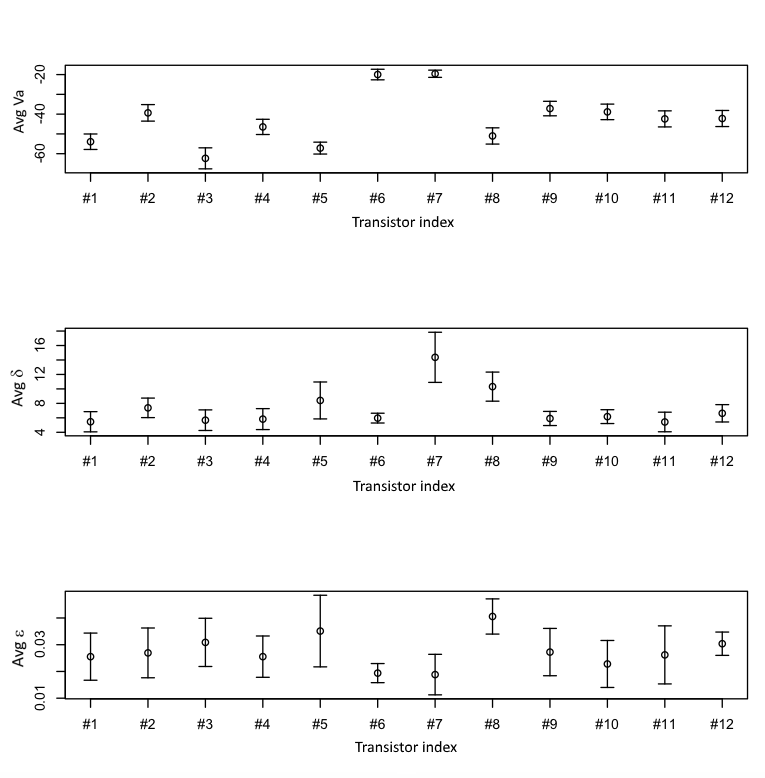} \\
  \caption{The estimated values (mean $\pm$ standard deviation along the scanning load lines) of $V_a$, $\delta$ (in percentage) and $\varepsilon$ for each of the considered BJTs. }
  ~\label{f:by_trans}
  \end{center}
\end{figure}

The linearity of such a relationship can be quantified by taking the mean squared error of the linear regression of the data $\theta \times I_b$ after $\theta$ has been normalized between 0 and 1.  This measurement is henceforth expressed as $\varepsilon$, being respective to a BJT and a given load line.  Relatively low values of $\varepsilon$ have been obtained (see Figure~\ref{f:by_trans}) for the considered BJTs, indicating that the relationship between the isolines angle $\theta$ and the base current is close to being linear in most cases.

The obtained results regarding the $V_a$ estimation, sharpness of intersections $\delta$, and errors $\varepsilon$ for each BJT are presented in Figure~\ref{f:by_trans}, in terms of the respective  mean $\pm$ standard deviation of the estimated values.  Such statistics are taken along the set of scanning load lines defined by $R_c$ and varying $V_{cc}$.  The coefficient of variation is henceforth presented in percentage,

The small error bars in the upper plot of Figure~\ref{f:by_trans} indicate that $V_a$ does not tend to vary much along the load lines.  Relatively small error bars are verified also for most cases in (b) and (c).  The sharpness of intersection values $\delta$ shown in (b) indicates that most of the considered BJTs can be modeled with similar quality, except for transistors $\#7$ and $\#8$, which yielded larger dispersions of intersections.  The rms errors $\varepsilon$ given in (c) indicate good overall fit of linear variation of $\theta$ with $I_b$ for all BJTs.  It is interesting to observe the similar measurements obtained in all cases for transistors $\#9$ to $\#10$, which are indeed from a same family.

\begin{figure}[]
  \begin{center}
  \includegraphics[width=\linewidth]{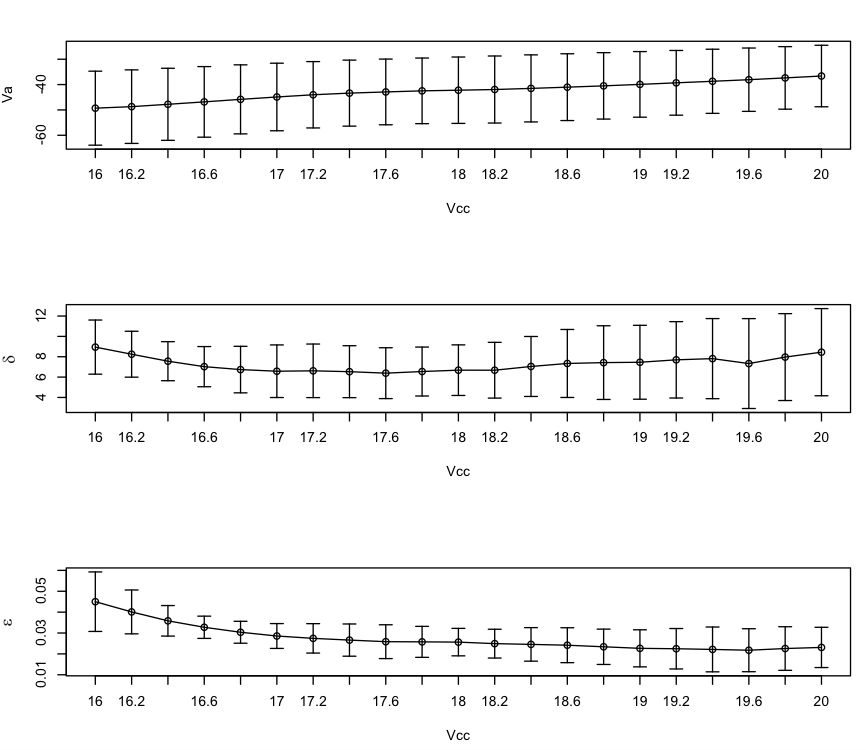} \\
  \caption{The estimated values (mean $\pm$ standard deviation for each scanning load line) of $V_a$, $\delta$ and $\varepsilon$ in terms of the values $V_{cc}$ defining the respective scanning load lines (with fixed $R_c = 800 \Omega$). }
  ~\label{f:by_Vcc}
  \end{center}
\end{figure}

The average $\pm$ standard deviation of the values o $V_a$, $\delta$ and $\varepsilon$, in terms of the of the values of $V_{cc}$ respective to the scanning lines, are depicted in Figure~\ref{f:by_Vcc}. These statistics were obtained for each $V_{cc}$, considering all 12 BJTs.  These results indicate some interesting features, such as the slight increase of estimated $V_a$ for larger values of $V_cc$.  At the same time, as $V_{cc}$ increases, the rms error $\varepsilon$ tends to be reduced.  Interestingly, the values of $\delta$, quantifying the goodness of fit of the Early model, tend to reach a minimum at about $V_{cc} = 17.8V$.  Further investigations are necessary in order to better understand such effects.

\subsection{Device Model Validation}

\begin{figure}[]
  \begin{center}
  \includegraphics[width=\linewidth]{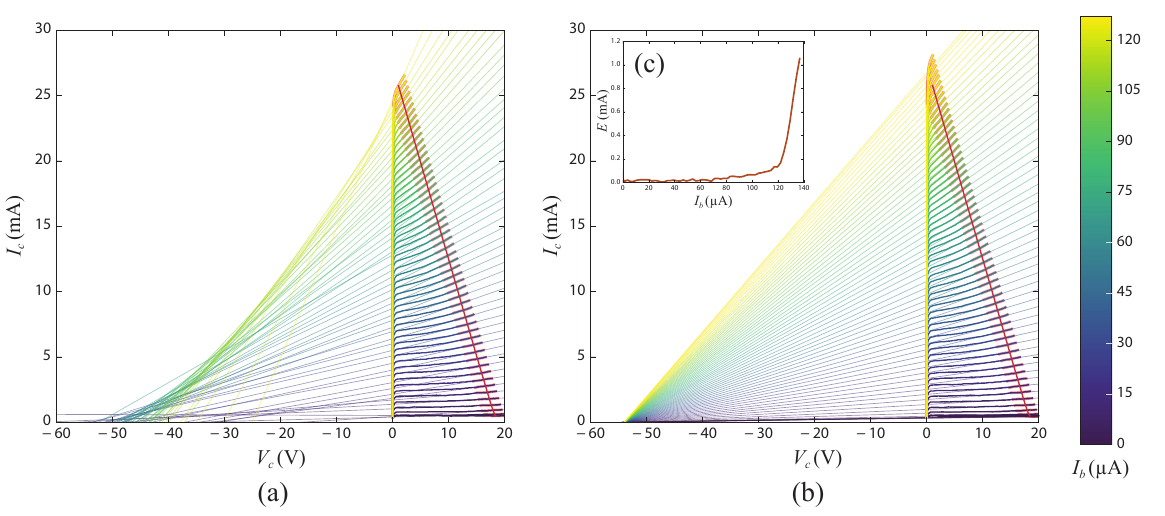} \\
  \caption{The original, experimental isolines obtained for transistor $\#1$ and respective prolongations (a).  The respective modeled isolines obtained from $V_a$ and intersections with the load line (b).  The respective regression errors $E$ in terms of $I_b$ (c). }
  ~\label{f:isoline_validation}
  \end{center}
\end{figure}

In order to validate the obtained Early models of the 12 considered BJTs, the following procedure was applied for each of them.  For a given BJT with Early voltage $V_a$, for each $I_b$, the respective modeled isoline corresponds to the straight line passing through the points $(-V_a, 0)$ and the intersection between the load line and the experimental isoline for that $I_b$.  Figure~\ref{f:isoline_validation}(a) shows the original experimental isolines for transistor $\#1$ and respective prolongations intersection the $V_c$ axis. Figure~\ref{f:isoline_validation}(b) depicts the several isolines obtained for transistor $\#1$ following the just described approach.  The regression errors are shown in Figure~\ref{f:isoline_validation}(c), with respect to each isoline and taking into account only the red region of the experimental data shown in Figure~\ref{f:isoline_validation}(b).  After a short interval of $I_b$, corresponding to the cutoff region of the BJT, the residues exhibit small values, which increase progressively until the saturation region, where large residue values are observed.  Similar results were obtained for all the other 11 BJTs considered in this work.  This result indicates a good concordance between the experimental and modeled isolines, supporting the relative accuracy of the proposed model for the considered devices.

\subsection{A Case Example:  Common Emitter Transfer Function}

\begin{figure}[]
  \begin{center}
  \includegraphics[width=0.9\linewidth]{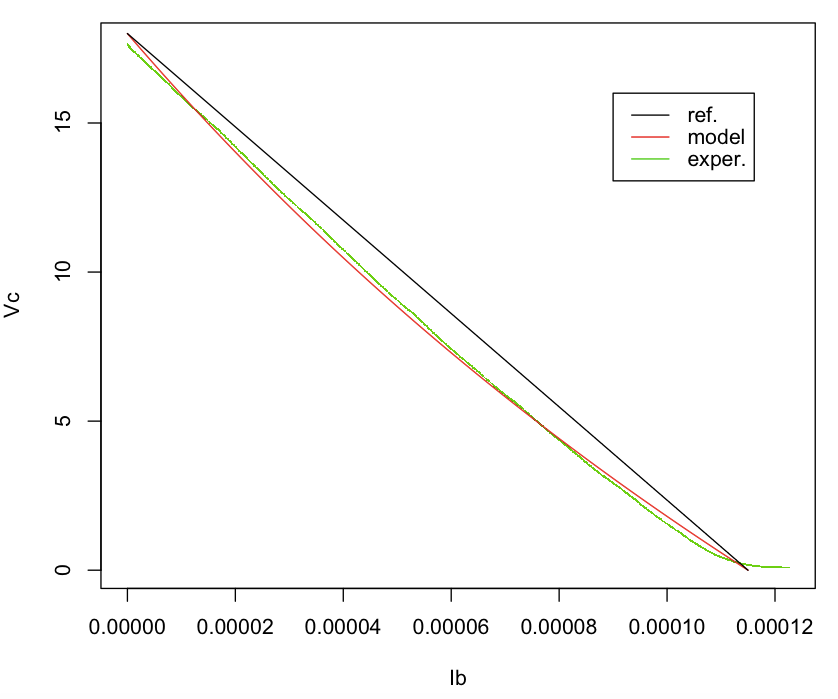} \\
  \caption{A case example illustrating the proposed Early modeling. In red, the transfer function obtained by using the model; the experimentally estimated transfer function is shown in green.  A reference, straight, transfer function is also included for reference. $R_c = 800 \Omega$ and $V_{cc}=18V$ for both modeled and experimental cases. }
  ~\label{f:case_example}
  \end{center}
\end{figure}

We conclude this article by presenting a complete example of application of the proposed Early model in order to design a common-emitter current amplifier, allowing a modeled transfer function to be contrasted with the respective one obtained from the experimental BJT data.  We chose BJT $\#1$ and make $R_c = 800 \Omega$ and $V_{cc}=18V$.   We know from the obtained results that $V_a \approx 53V$ for this transistor and parameters setting.  Thus, by substituting this value in Equation~\ref{eq:Vcth}, we immediately have the voltage load line shown in red in Figure~\ref{f:case_example}.  The experimental transfer function obtained for this same BJT under the same parameters is shown in green in this figure.   A reference, straight transfer function is also included in the figure.  It is clear that an impressive agreement between model and real voltage transfer functions has been achieved, which is surprising given the seeming complexity of the BJT properties, including variation of $\theta$ with $I_b$.  Such a result illustrates the potential of the suggested methodology for modeling and predicting the properties of circuit configurations for a given, specific BJT.  It is observed that larger deviations can be, in principle, expected for other values of $R_c$ and $V_{cc}$ because of the observed relatively small variations of $V_a$ in terms of these two parameters.  Further investigations are required to show more systematically to what an extent such transfer function models extend to other circuit configurations and devices/types.

\section{THD Characterization in Terms of the Early Voltage}

\begin{figure}[]
  \begin{center}
  \includegraphics[width=\linewidth]{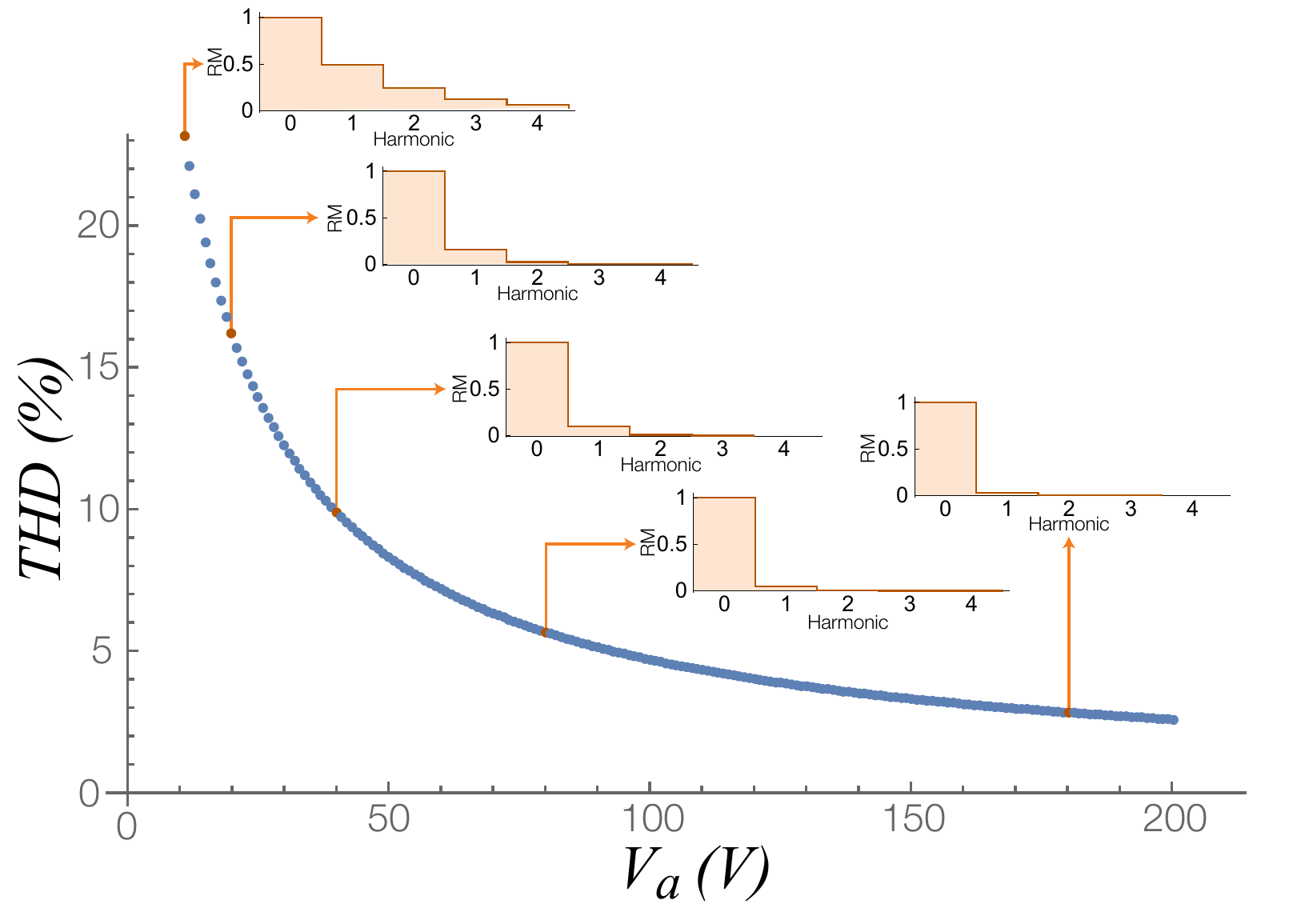} \\
  \caption{The THD values obtained from the Early model considering $R_C = 900 \Omega$ and $V_{cc} = 25V$, as well as the relative magnitude, RM, of harmonics for several $V_a$ values.}
  ~\label{f:thd}
  \end{center}
\end{figure}

The linearity of a transfer function can be explored by obtaining its THD.  We performed such an analysis by using transfer functions obtained by the proposed Early model.   More specifically, we used $R_C = 900 \Omega$ and $V_{cc} = 25V$, and considered several equispaced values of $V_a$ between $10V$ and $200V$.  Figure~\ref{f:thd} shows the resulting THD values in terms of $V_a$.  This figure also shows, for several values of $V_a$ the relative magnitude of the Fourier harmonics.  It is clear from these results that the THD tends to decrease steadily with $V_a$, reaching the relatively high value of $8\%$ for the typical Early voltage of $50V$ typically found in real-world BJTs.  The harmonic distribution tends to progressively become concentrated in the first harmonic component.

\section{Concluding Remarks}

Bipolar transistors, in their discrete or integrated implementations, have constituted the basis of modern electronics, allowing outstanding advances in computing, networking and control.  In linear electronics, as in many other scientific and technological approaches, \emph{linear operation} has been eagerly sought because of its ability to control the intensity of signals while not adding distortions.  Unfortunately, no fully linear device or circuit can be found in the real-world.  A good deal of such non-linearity stems from the own adopted devices, which often correspond to BJTs.  The intrinsic non-linearity of such devices, compounded by their varying properties, have provided one of the biggest challenges to their widespread application, which was ultimately only achieved by using negative feedback~\cite{chen2016active}.  In a recent work~\cite{costa2016negative}, it has been shown that the variability of small signal BJTs of a same type may not so great as usually thought.  Indeed, different types of small signal BJTs have their properties so consistent as to allow good discrimination and clustering.  At the same time, it was also found under certain circumstances~\cite{costa2016negative} that negative feedback, even when intense, may not be capable of completely achieving device invariance, while contributing moderately to improving the linearity but also increasing substantially the output resistance in the considered circuit configurations.  For all such reasons, it seems to make sense to consider the specific properties of the adopted devices during the design, implementation and characterization of circuits in linear electronics.  Such an approach requires good models of BJTs.

The current work addressed such an issue by proposing a simple, intuitive and geometrical model of BJT operation that can be used to model and characterized devices and respective circuit configurations.  This has been achieved by taking into account the underlying geometrical organization of the characteristic surfaces of such devices as implied by the Early effect/voltage formulation.  Typically, models of BJTs are limited to current-equispaced horizontal parallel lines in the $I_c \times V_c$ space, not incorporating the Early voltage.   The proposed model, though not considering the saturation or cut-off regimes, is derived precisely by taking into account this term, which yields a simple and intuitive organization of the BJT characteristic surface in terms of a beam of angle-equispaced isolines that intersect at a point $(-V_a,0)$ along the $V_c$ axis, where $V_a$ is the Early voltage.  Though the model can immediately be extended to account for more general relationships between the angles $\theta$ of the isolines and the base current $I_b$, the experimental results reported in this worked suggest that, in practice and for the considered devices, such a variation can be approximated by a linear relationship, which gives rise to a purely geometrical version of the Early model, with the isoline angles being proportional to the base current $I_b$.  In addition to providing a simple and yet relatively complete model of BJTs in the linear regime, the proposed approach also presents good potential to be used as a didactic aid in physics and electronics programs.

Several have been the results achieved by using the proposed geometrical model.   First, it clarifies the main reasons potentially conspiring to undermine the linearity of a BJT, namely:  (a) the value of $V_a$, with larger values contributing to the linearity; and (b) the way in which the isoline angles $\theta$ varies with $V_b$ and $I_b$.  These two sources of distortion, are likely independent one another, and can be properly accounted for by the Early model.  There is a third potential cause of non-linearity, which is implied by deviations of the isolines from being fully straight.   This possibility led to the introduction of a \emph{local} version of the Early model, where only portions of the isolines (limited by a region of interest) are taken into account for the derivation of $V_a$.  Such a local approach has been used to obtain estimations of $V_a$ and related parameters by scanning the isolines for successive values of $V_{cc}$, while keeping the value of $R_c$ fixed.  It should be observed that other scanning approaches can be considered, such as varying $R_c$ while keeping $V_{cc}$ fixed, etc.

Though interesting by itself, the usefulness of the proposed Early model depends on how well its basic hypotheses are verified in practice.   In order to probe into this important issue, we used 12 small signal BJTs of different types to experimentally derive sets of isolines to be used for the Early approach validation.  We have also used the coefficient of variation of the intersections with the $V_c$ axis, in order to quantify the sharpness of that intersection (and, henceforth, provide some insight about the quality of $V_a$ estimation), and the mean squared error of the linear regression between $\theta$ and $I_b$.

First, we considered several regions of interest, by varying $R_c$ for a fixed $V_{cc}$ value.  Interestingly, the sharpness of the intersection tended to vary moderately with $R_c$, achieving its best value for  $R_c \approx 800 \Omega$ for the considered samples.  This value of $R_c$ was adopted for the subsequent experiments.  Also, such a variation indicates that the isolines are not completely straight, confirming their visual appearance, and suggesting additional research aimed at characterizing such deviations, so that they can be eventually incorporated into the Early model.  Yet, to a large extent, the $V_a$ values obtained for each scanning load line, for each BJT, resulted mostly stable and consistent for $R_c \approx 800 \Omega$, as indicated by relatively small values of $\delta$.  Contrariwise, the Early voltages estimated for each BJT often resulted markedly different, indicating distinct properties of each considered device/type. A THD analysis of transfer functions derived from the Early approach was also developed and discussed, revealing relatively high THD values for typical Early voltage of $50V$. These results illustrate the potential of using the Early model to reflect such specific device variations during circuit design and characterization, especially when no or little negative feedback is incorporated.  

Another important finding is the fact that $\theta$ tends to exhibit a mostly linear variation with $I_b$.  This phenomenon, which needs to be further investigated, allows the proposed Early model to become particularly simple, fully geometrical, with the isolines being directly characterized by their respective angles $\theta$.  This contributes to the simplicity and practicality of the proposed approach, being inherently compatible with the current-equispaced isolines model commonly used.   Also important, from both theoretical and practical points of view, is the fact that $V_a$ tends do decrease with $V_{cc}$, while the linearity of the relationship $\theta \times I_b$ increases.  Further investigations are necessary to understand such phenomena further, as well as to verify the extensibility and accuracy of the Early approach to other circuit configurations and devices (samples or types), which may also include MOS and FET transistors, as well as their embedding in integrated circuit designs.

All in all, we have shown that the Early voltage provides an important subsidy for defining the geometrical organization of BJT operation alone or in circuit configurations, not taking into account the saturation and cut-off regimes and transitions.  Moreover, the experimental verification, at least for the considered BJTs, that the isolines angles tend to vary linearly with $I_b$ bears an important implication in keeping the Early model simple in the sense that the isolines angles are directly proportional to $I_b$.  This interesting property, as well as a more systematic characterization of the Early geometry implications to non-linearity of the transfer functions, will be investigated further in  subsequent works.  Ultimately, the Early approach and respectively obtained experimental results suggest that, despite their seeming complexity, BJTs are to a large extent simple devices, in the sense that their properties can be mostly represented in terms of just the Early voltage, plus eventually the speed $s$ of how the isoline angles vary with $I_b$.  Such a phenomenon accounts for the fact that our previous study~\cite{costa2016negative} identified only 1 or 2 statistical degrees of liberty (through principal component analysis) in the dispersion of several properties of 60 real world BJTs.  It should be also observed that the two parameters in the Early model, namely the Early voltage $V_a$ and the constant $s$, provide an effective indexing system that can be used to systematically classify different types of transistors in terms of their properties.  

\section*{Acknowledgements}
L. da F. Costa thanks CNPq (Grant no. 307333/2013-2) for support. F. N. Silva acknowledges FAPESP (Grant No. 15/08003-4). C. H. Comin thanks FAPESP (Grant No. 15/18942-8) for financial support. This work has been supported also by FAPESP grant 11/50761-2.

\bibliographystyle{spmpsci}      
\bibliography{references}   

\end{document}